\documentclass[12pt, draftclsnofoot, onecolumn]{IEEEtran}
\usepackage[margin=1in]{geometry}
\usepackage{fix-cm}
\usepackage{cite}
\usepackage{amsmath,amssymb,amsfonts}
\usepackage{algorithmic}
\usepackage{graphicx,color}
\usepackage{textcomp}
\usepackage{hyperref}
\usepackage{balance}
\usepackage{booktabs}
\usepackage{amsfonts} % For \mathbb
\usepackage{bm} % For bold math symbols
\usepackage{graphicx}
\usepackage{lipsum}
\usepackage{comment}
\usepackage{caption}
\usepackage{subcaption}
\usepackage{tabularx}
\usepackage{booktabs}
\usepackage{float}        % for precise figure placement, if needed
\usepackage{textcomp}
\usepackage{authblk} % For affiliations
\usepackage{placeins}  % Add this in the preamble
\usepackage{multirow} 
\usepackage[font=small,compatibility=false]{caption}

\def\BibTeX{{\rm B\kern-.05em{\sc i\kern-.025em b}\kern-.08em
    T\kern-.1667em\lower.7ex\hbox{E}\kern-.125emX}}
\AtBeginDocument{\definecolor{ojcolor}{cmyk}{0.93,0.59,0.15,0.02}}

\title{Collection: UAV-Based RSS Measurements from the AFAR Challenge in Digital Twin and Real-World Environments}
\author[1]{Saad Masrur}
\author[1]{\"Ozg\"ur \"Ozdemir}
\author[1]{An\i l G\"urses}
\author[1]{\.{I}smail G\"uven\c{c}}
\author[1]{Mihail L. Sichitiu}
\author[1]{Rudra Dutta}
\author[1]{Magreth Mushi}
\author[1]{Thomas Zajkowski}
\author[1]{Cole Dickerson}
\author[1]{Gautham Reddy}
\author[1]{Sergio Vargas Villar}
\author[1]{Chau-Wai Wong}
\author[1]{Baisakhi Chatterjee}
\author[1]{Sonali Chaudhari}
\author[1]{Zhizhen Li}
\author[1]{Yuchen Liu}
\author[2]{Paul Kudyba}
\author[2]{Haijian Sun}
\author[3]{Jaya Sravani Mandapaka}
\author[3]{Kamesh Namuduri}
\author[4]{Weijie Wang}
\author[4]{Fraida Fund}

\affil[1]{North Carolina State University, Raleigh, NC, USA}
\affil[2]{University of Georgia, Athens, GA, USA}
\affil[3]{University of North Texas, Denton, TX, USA}
\affil[4]{New York University, Brooklyn, NY, USA}

\begin{document}
\maketitle

\vspace{-0.9cm}
\begin{abstract}
This paper presents a comprehensive real-world and Digital Twin (DT) dataset collected as part of the AERPAW Find A Rover (AFAR) Challenge, organized by the NSF Aerial Experimentation and Research Platform for Advanced Wireless (AERPAW) testbed and hosted at the Lake Wheeler Field in Raleigh, North Carolina. The AFAR Challenge was a competition involving five finalist university teams, focused on promoting innovation in unmanned aerial vehicle (UAV)-assisted radio frequency (RF) source localization. Participating teams were tasked with designing UAV flight trajectories and localization algorithms to detect the position of a hidden unmanned ground vehicle (UGV), also referred to as a rover, emitting probe signals generated by GNU Radio. The competition was structured to evaluate solutions in a DT environment first, followed by deployment and testing in AERPAW\textquotesingle s outdoor wireless testbed. For each team, the UGV was placed at three different positions, resulting in a total of 29 datasets—15 collected in a DT simulation environment and 14 in a physical outdoor testbed. Each dataset contains time-synchronized measurements of received signal strength (RSS), received signal quality (RSQ), GPS coordinates, UAV velocity, and UAV orientation (roll, pitch, and yaw). Data is organized into structured folders by team, environment (DT and real-world), and UGV location. The dataset supports research in UAV-assisted RF source localization, air-to-ground (A2G) wireless propagation modeling, trajectory optimization, signal prediction, autonomous navigation, and DT validation. With 300k time-synchronized samples from the real-world experiments, the AFAR dataset enables effective training/testing of deep learning (DL) models and supports robust, real-world UAV-based wireless communication and sensing research.\\

 \noindent\textbf{IEEE SOCIETY/COUNCIL:} 
IEEE Communications Society (ComSoc), IEEE Vehicular Technology Society (VTS), IEEE Aerospace and Electronic Systems Society (AESS)  

\vspace{0.5em}
\noindent\textbf{DATA DOI/PID:} 
\href{https://doi.org/10.5061/dryad.18931zd4g}{10.5061/dryad.18931zd4g}  

\vspace{0.5em}
\noindent\textbf{DATA TYPE/LOCATION:} 
Time-series; AERPAW Lake Wheeler Field Labs, Raleigh, North Carolina, USA; AERPAW Digital Twin for Lake Wheeler Field Labs

\end{abstract}


% \maketitle
\vspace{1.2cm}
\section*{BACKGROUND} 

% \textcolor{blue}{Guide Lines to write background section: Background must provide an overview of the data collected and discuss how it fits with other comparable, published datasets. Authors must make clear the data's value and how it can be reused. Authors must also summarize any previous publication made using this data, with a brief summary and citation for each time used. \textbf{DO NOT} include a paragraph on how your article is organized; all articles of this type are organized the same way.}

\begin{figure}[t!]
    \vspace{-0.4cm}

    \centering
    \includegraphics[width=0.75\linewidth]{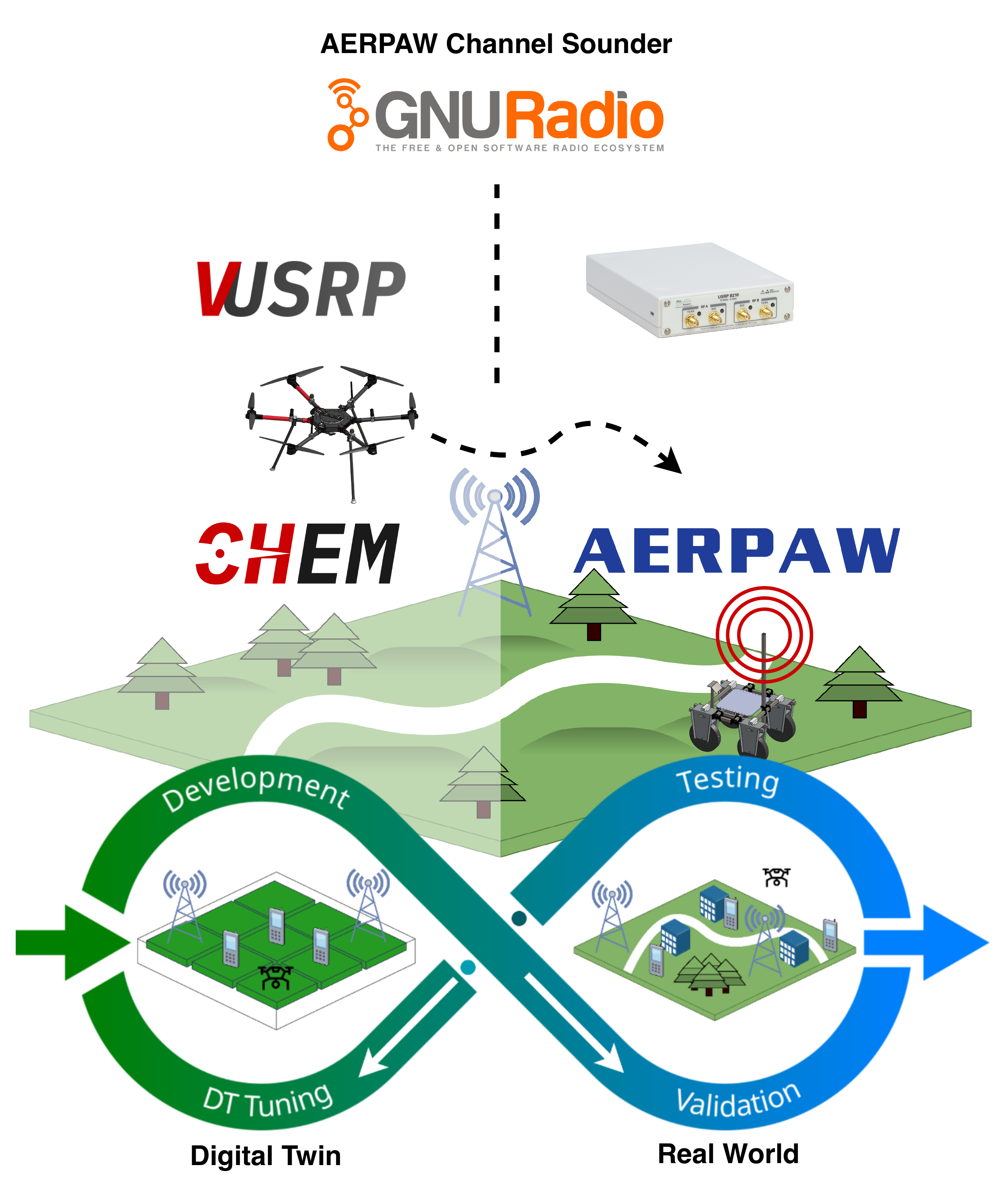}
    \caption{AFAR challenge on AERPAW digital twin and real world.}
    \label{fig:afar_system}
    \vspace{-0.5cm}
\end{figure}

Unmanned aerial vehicles (UAVs) are integral to location-based applications such as search and rescue, surveillance, and radio frequency (RF) source localization due to their autonomous 3D navigation, high-altitude signal reception, and superior mobility compared to terrestrial networks. However, most research involving UAVs for assisting rescue operations, localization, or advancing wireless communication is conducted in simulated environments that do not fully capture real-world complexities, leading to challenges in translating solutions to practical deployments. Accurate algorithm development for UAV-assisted applications must account for real-world channel dynamics, mobility variations, and environmental effects—factors often overlooked in simulation-based studies. However, the lack of testbed facilities limits comprehensive validation and testing in a real-world environment. To bridge this gap, there is a growing need for integrated platforms that support both virtual development environments and real-world testing. These platforms enable researchers to design, implement, and iteratively refine solutions in controlled digital settings before transitioning to field deployment. The Aerial Experimentation and Research Platform for Advanced Wireless (AERPAW) addresses this need by offering a digital twin (DT) for virtual experimentation and a physical outdoor testbed for real-world deployment. As illustrated in Fig.~\ref{fig:afar_system}, AERPAW facilitates seamless transition from simulation to deployment, supporting scalable and repeatable evaluation of UAV-assisted wireless systems. The aforementioned figure is adapted with modifications from \cite{gurses2024digital} to highlight the differences between the DT and real-world environments.

Achieving high-accuracy solutions for UAV-assisted applications requires rigorous data-driven analysis and the collection of real-world wireless communication data under realistic flight and mobility conditions to effectively model the dynamic nature of wireless channels, environmental variations, and system constraints. UAV-assisted datasets offer the opportunity to capture how signal behavior evolves with altitude, orientation, and environmental dynamics. To enable comprehensive modeling and validation, the collected dataset should include key wireless performance metrics such as Received Signal Strength (RSS) and Received Signal Quality (RSQ), along with precise timestamps to support temporal feature extraction and time-series analysis of signal behavior over the UAV’s flight. Additionally, it should contain the GPS coordinates of the UAV to allow accurate spatial correlation between signal measurements and UAV position.

\vspace{-0.2cm}
\subsection*{AFAR Competition} 
\begin{figure}
% \vspace{-0.5cm}
\includegraphics[width=0.9\linewidth]{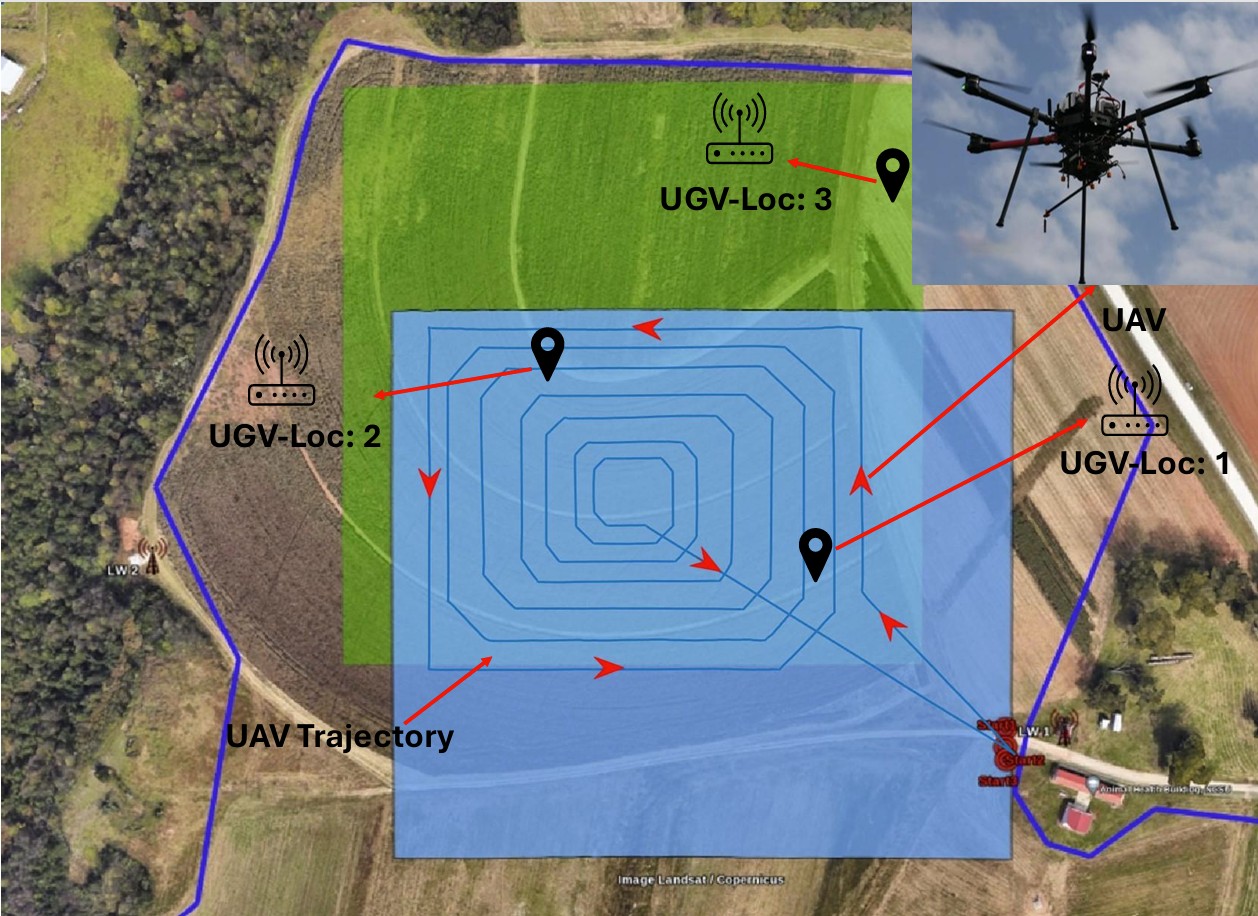}
	\centering
% \vspace{-0.1cm}
	\caption{Experimental setup of the AFAR Challenge, illustrating the UAV flight zone (blue) and the region where the UGV can be positioned (green), with the UAV used in the experiment depicted in the top right corner. }
	\label{fig1}
\vspace{-0.6cm}
\end{figure}

AERPAW hosted the AERPAW Find A Rover (AFAR) Challenge, inviting students from universities across the United States to bridge the gap between research and practical implementation in UAV-assisted RF source localization. The challenge tasked participants with designing and developing innovative algorithms to localize an unmanned ground vehicle (UGV), also referred to as a rover, using a UAV-based receiver. The UAV, a custom hexacopter shown in the upper-right corner of Fig.~\ref{fig1}, collects RF signals during flight, while the UGV serves as the RF source transmitter. The resulting dataset includes time-synchronized measurements such as RSS, RSQ, GPS coordinates, and UAV orientation data, making it a valuable resource for spatiotemporal analysis and algorithm validation.

The experimental setup allowed the UGV to be placed anywhere within a designated area (marked in green in Fig.~\ref{fig1}), while the UAV was restricted to flying within a predefined flight zone (marked in blue). A sample UAV trajectory is overlaid within the flight zone. Notably, portions of the green area fall outside the blue region, forming designated no-fly zones. These constraints prevent the UAV from flying directly above certain UGV positions, thereby making the localization problem more challenging, necessitating advanced localization strategies along with optimized trajectory planning. Teams were allowed to develop either an autonomous flight trajectory or a predefined waypoint-based trajectory for the UAV. The UAV was permitted to fly at altitudes ranging from 20~m to 110~m, with a maximum speed of 10~m/s. The real-world measurements were collected at the Lake Wheeler Field, a rural outdoor test site that includes large open grass fields, surrounding tree lines, scattered bushes, and limited man-made structures, as shown in Fig.~\ref{fig1}.

\subsection*{Scoring and Ranking}

Two key metrics were used to evaluate  performance:  
\begin{itemize}
    \item \textbf{Fast Localization Accuracy (FLA)}: Requires teams to estimate and submit the UGV's location using only data collected within the first 3 minutes of UAV flight.
    \item \textbf{Long-Term Localization Accuracy (LTLA)}: Assesses the final localization accuracy based on all measurements gathered over the full 10-minute UAV flight, with the final estimate submitted at the end of the mission.
\end{itemize}

In the first round, teams developed their localization algorithms within AERPAW’s virtual environment, leveraging its DT \cite{gurses2024digital}, where their solutions were tested for three different UGV locations, as shown in Fig.~\ref{fig1}, with the UGV position hidden by the organizers during the competition. The ranking of teams was based on the root mean square error (RMSE) of their location estimates across these scenarios. Based on DT performance, the top five teams out of the nine registered teams advanced to the next phase, where their solutions were deployed and tested in AERPAW’s physical testbed at Lake Wheeler, North Carolina. Evaluations are performed both in the DT and in the testbed. The performance in the DT did not influence the final ranking of the teams was determined solely by the score achieved during the real testbed run. Nevertheless, results from the DT environment were included alongside real-world testbed data to facilitate direct comparison and analysis of the differences between emulated and physical experiments.
\subsection*{Dataset Overview and Uniqueness}
The AFAR Challenge dataset contains data collected from the top five teams that participated in the competition, with each team developing distinct UAV flight trajectories and localization algorithms to estimate the position of a UGV. The dataset includes both real-world and DT data, consistently structured across both environments. For each team, the UGV was deployed at three different locations (Loc-1, Loc-2, Loc-3), resulting in a total of 30 datasets, with 14 from real-world experiments and the remaining 15 from the DT environment. The participating teams were Team SunLab from the University of Georgia (T-288), Team NYU Wireless from New York University (T-300), Team Eagles from the University of North Texas (T-301), Team Wolfpack from NC State University (T-309), and Team Daedalic Wings from NC State University (T-328).

To the best of the authors' knowledge, no prior challenge of this nature has been organized, nor has such a dataset been publicly released. This initiative represents the first large-scale dataset designed to facilitate advanced wireless communication analysis using UAVs, with a particular focus on the RF localization and air-to-ground (A2G) channel characterization. While RF source localization has been explored in previous studies, it has predominantly been investigated in simulated environments \cite{kwon2023rf,8467555,yang2022impact}. Some experimental efforts have focused on time difference of arrival-based localization techniques, as in \cite{dickerson2025tdoa}, which examines the impact of altitude, bandwidth, and non-line-of-sight (NLOS) bias on 3D UAV localization. Meanwhile, the studies in \cite{lyu2024fixed} and \cite{aerpaw_a2g_gurses} examine A2G channel characteristics through experimental measurements conducted in a rural environment, where recorded signal data was later used to refine path loss models. However, this dataset \cite{lyu2024fixed} is not publicly available, and the study primarily considers a single, fixed-location RF source for channel characterization rather than localization. In contrast, our dataset uses a stationary RF source placed at multiple distinct locations across trials, enabling analysis and benchmarking of UAV-based RF localization algorithms.

% However, this dataset is not publicly available, and the study primarily considers a static RF source at a fixed location, focusing on channel characterization rather than localization. 

% \textcolor{red}{If the authors are aware of any publicly available datasets similar to ours, they should provide a brief literature review in this section.}

% The AFAR dataset comprises 14 distinct datasets collected from five different teams, where the UGV was positioned at three different locations while each team independently designed and executed UAV flight trajectories to collect RF signal data (details provided in Section IV). The dataset has been publicly released via Dryad and can be accessed using the DOI: https://doi.org/10.5061/dryad.18931zd4g.
\vspace{-0.2cm}
\subsection*{How the Dataset can be Reused}
This dataset plays a pivotal role in advancing UAV-assisted RF source localization by enabling researchers to design, train, and benchmark algorithms under diverse flight paths and propagation conditions. In addition, it supports A2G channel modeling by capturing key characteristics such as propagation effects, fading, noise variations, and the influence of UAV speed and orientation. Together, these features provide critical insights into UAV-assisted wireless communication systems, making the dataset a valuable resource for a wide range of research domains. The key contributions of the AFAR dataset are as follows.

\noindent \textbf{\textit{Performance Analysis.}} The dataset enables a comprehensive performance analysis of UAV-assisted RF source localization. Researchers can evaluate how trajectory design, speed variations, and algorithm choice affect localization accuracy and efficiency.

\noindent \textbf{\textit{Data-driven real-world applications.} } Unlike purely simulated datasets, the AFAR dataset consists of real-world measurements, making it highly valuable for deep learning (DL) applications. It provides an empirical foundation for training DL models in various domains, including RF localization, UAV navigation, tracking, and autonomous path planning. Additionally, researchers can use this dataset to validate theoretical models and AI-driven approaches, ensuring practical viability. This bridge between theoretical advancements and experimental validation accelerates the translation of research innovations into deployable UAV solutions, fostering advancements in agriculture, disaster management, and aerial sensing applications.

\noindent \textbf{\textit{A2G channel propagation modeling.}}  A2G propagation plays a critical role in UAV-assisted wireless communication, directly impacting link reliability, coverage, and system performance. Accurate modeling of A2G characteristics is essential for enabling key UAV applications such as RF source localization, aerial surveillance, emergency response, edge computing, and beyond 5G (B5G) aerial connectivity. Unlike terrestrial links, A2G channels exhibit unique propagation characteristics influenced by factors such as elevation angle, UAV altitude, orientation (roll, pitch, and yaw), speed, and environmental obstructions. These A2G links are influenced not only by distance-dependent path loss but also by diffraction, shadowing, and multipath effects, necessitating specialized models beyond traditional approaches. Additionally, shadowing effects caused by terrain, vegetation, and urban structures, as well as UAV-specific challenges such as antenna misalignment and signal obstruction due to the UAV body, significantly impact communication performance. The AFAR dataset facilitates the development of accurate A2G channel models, enhancing UAV adaptability across diverse operational scenarios.

The dataset collected during the challenge was utilized by participating teams to develop and evaluate their localization algorithms. The AFAR dataset enabled teams to assess the performance of their approaches under practical conditions. A detailed analysis of the localization methodologies employed by the top three teams, along with their utilization of the AFAR dataset, is provided in \cite{kudyba2024uav}, which also offers an overview of the challenge framework. Additionally, another participating team documented their approach in \cite{10619824}, while the implementation details of the third-ranking team are available in \cite{10757589}.
The study in \cite{masrur2025bridging, rahman2025characterization} leveraged the AFAR dataset to bridge the gap between simulation and real-world deployment. The authors in \cite{masrur2025bridging} proposed an enhanced two-ray path loss model to improve A2G channel modeling by incorporating shadowing effects induced by the UAV body. To accurately capture these effects, they utilized the roll, yaw, and pitch angles provided in the dataset. The AFAR dataset was also used to validate a DL-based localization model trained on simulated data, demonstrating its effectiveness in real-world conditions.
% Furthermore, the authors trained a DL-based localization model on a simulated dataset and subsequently validated its performance using the real-world measurements from the AFAR dataset, ensuring robustness and adaptability in practical scenarios.

% This dataset is essential in modeling the A2G propagation model, effect of fading, noise, channel coherence time, the effect of speed on the signal power, and the shadowing. shadowing happens due to the UAV body, antenna gains effect, and other factors.  

% Datagented during the AFAR channels is the first big dataste available which can provide insights about air to ground (A2G) channel models and help researcher to get a sense or real life environment. 
\vspace{-0.2cm}
\section*{COLLECTION METHODS AND DESIGN} 

% \textcolor{blue}{Generic Guideline for this section: Collection Methods and Design must provide details on how that data was collected. This includes details on any hardware/system designs used to collect the data (i.e., data acquisition). In addition, the steps and procedures used to collect and process the data in its final form (i.e., computational processing). It is recommended that the authors provide diagrams that show the overall system/procedure used.}

% \textcolor{red}{It would be great if Dr. Ozgur and Anil could write this section, ensuring it spans at least half a page or more and includes at least one diagram. Dr. Ozgur and Anil can explain the working mechanism of the channel sounder, how data is collected, how this data is made available in the DT in real time.  Including one diagram is necessary as per the requirement of the paper's format. Please let me know if you need any assistance with this.}

% \begin{figure}[htbp]
%     \centering
%     \includegraphics[width=\linewidth]{Figures/AFAR_System.pdf}
%     \caption{AFAR challenge on AERPAW digital twin and real world.}
%     \label{fig:afar_system}
%     \vspace{-0.3cm}
% \end{figure}
The data collection was carried out using the AERPAW platform \cite{aerpaw_portal}, an advanced wireless testbed that supports experimentation with programmable radios and autonomous vehicles in both simulated and real-world environments. Teams initially developed and evaluated their localization solutions in AERPAW’s DT environment, which allows remote testing and refinement. Once finalized, these solutions were executed in the AERPAW’s physical testbed.

Within the platform's DT environment, each team initiates its experiment by choosing two nodes: one UGV, also known as the rover, and one UAV.
In this setup, the UGV runs the transmitter part of the AERPAW channel sounder sample experiment \cite{aerpaw_ge2}, while the UAV runs the receiver part. UAV does not know the location of the rover, and it tries to detect its position based solely on the RSS measurements at the UAV.
% Participants get themselves familiar with the AFAR challenge by first trying the \textit{AERPAW to Find a Rover Sample Code} \cite{aerpaw_ge2}. In the sample code, the UAV moves forward as long as the RSS improves; if the signal strength decreases, it turns 90 degrees to the right. Although this simple algorithm does not guarantee accurate location of the rover, it introduces participants to key concepts: how to read RSS measurements and how to control the movement of the UAV within geofence restrictions imposed by AERPAW.
As the UAV approaches the rover, RSS generally increases. However, due to the nature of wireless channel variations and the influence of antenna patterns, the RSS does not always increase monotonically. This variation in signal strength poses a challenge in designing robust localization algorithms.
Participants develop their strategies in the AERPAW DT—an environment that emulates the behavior of the real-world testbed. While the DT's channel characteristics do not exactly mirror reality, AERPAW supplements this limitation by providing real testbed data. These datasets include RSS as a function of distance between the transmitter and the receiver, giving teams valuable insight into real-world propagation behavior. Once all teams finalize their algorithms, the code is first tested in the DT, and the same code (in software containers) is moved as it is to the physical testbed. The evaluation in the DT also helped AERPAW operators to identify and eliminate potential bugs before running the experiment in the testbed.

% \textcolor{red}{Once all teams finalize their algorithms, the code is first tested within the digital twin. For the AFAR challenge, three locations are selected for the rover, and the code of participating teams is evaluated based on the same locations. Evaluations were performed both in the digital twin and in the testbed. The performance in the digital twin did not influence the final ranking of the teams. The final ranking was determined solely by the score achieved during the real testbed run. Nevertheless, the digital twin results were also included in addition to the testbed results as part of this dataset.}

The channel sounder is implemented using open-source software, GNU Radio Companion. At the transmitter, a degree-12 Galois Linear Feedback Shift Register generates a pseudo-random bit sequence with a period of 4095 bits. This sequence was interpolated by a factor of 16 and passed through a root-raised cosine filter before being transmitted by a USRP device at a sampling rate of 2 MHz. In the DT, IQ samples generated by the channel sounder transmitter are sent to a virtual USRP (V-USRP). AERPAW experiments consist of containers called virtual machines (VMs). The experimenter code runs in Experiment VM (E-VM).

%\textcolor{red}{Dr. Ozgur, can you please mention which antenna was used, like brand or something? After including all of it, can you please rewrite these few lines: In the testbed setup, USRP B205mini devices were used, connected to RF front-ends and antennas, and an Intel NUC 10 (Intel i7-5550U with 64 GB RAM and 1 TB SSD) It is used as a companion computer for UAV, and a 1 W wide-band low-noise power supply amplifier with filters is used at the front ends of the USRP. }.

In the DT, the E-VM is connected to a channel emulator VM called CHEM-VM. The V-USRP inside the E-VM sends IQ samples to the CHEM-VM. Based on the relative positions of the UAV and UGV, CHEM-VM applies realistic channel effects before forwarding the data to the UAV’s receiver E-VM. In the real testbed, the containers from the DT are moved to the computer inside the portable nodes attached to the UAV and the UGV. As an example, the same channel sounder transmitter runs on a real USRP, and this time, the IQ samples are up-converted to RF, which is radiated through antennas, propagating through the physical wireless channel. 

The portable nodes at the UGV and the UAV contain a USRP B205mini, a compact, full-duplex software-defined radio that supports operation over a frequency range of 70~MHz to 6~GHz. The USRP B205mini is connected to an Intel NUC 10 with an i7-10710U processor, 64~GB of RAM, and a 1~TB SSD, which provides the computational resources necessary for real-time signal processing. 
The transmit path begins with the output from the USRP B205mini, which is passed through a band-pass filter (VBFZ-3590-S+) operating in the 3.0~GHz to 4.2~GHz range. This filter suppresses unwanted spectral components. The filtered signal is then amplified using a power amplifier (ZVE-8G+) that operates from 2~GHz to 8~GHz and provides up to 1~Watt of output power. Finally, the signal is transmitted via a wideband antenna (SA 1400-5900) with coverage from 1.4~GHz to 5.9~GHz.
The receive path shares the same antenna, enabling bidirectional operation and reducing hardware complexity. The incoming signal is first filtered using the same band-pass filter (VBFZ-3590-S+) to remove out-of-band interference. The filtered signal is then amplified by a low-noise amplifier (ZX60-83LN-S+) to improve the signal-to-noise ratio before being passed to the USRP B205mini for digitization and further processing.

At the UAV, the channel sounder receiver code is implemented. The receiver first compensates for the frequency offset between the transmitter and the receiver. Then, the compensated signal is correlated with the original 4095-bit sequence used at the transmitter. This correlation produces a Channel Impulse Response (CIR), where the power of the CIR peak is measured and logged as power and used by the UAV to determine the location of the rover. The RSQ log generated by the channel sounder is the difference between the peak power and the average power in the CIR outside the peak location.

\begin{table}[!t]  %H
\vspace{-0.5cm}
\caption{Statistics of RSS, signal RSQ, and UAV speed collected at Loc-1, Loc-2, and Loc-3 for team-288.}
\centering
\begin{tabular}{|l|l|c|c|c|}
\hline
\textbf{Feature} & \textbf{Stat.} & \textbf{Loc-1} & \textbf{Loc-2} & \textbf{Loc-3} \\ \hline

\multirow{4}{*}{\textbf{RSS (dB)}} 
& Mean & 14.42 & -7.248 & -19.95 \\
& Std & 13.99 & 25.85 & 15.06 \\
& Max & 31.60 & 29.61 & 13.15 \\
& Min & -78.64 & -63.90 & -64.85 \\ \hline

\multirow{4}{*}{\textbf{RSQ (dB)}} 
& Mean & 69.87 & 59.12 & 53.06 \\
& Std & 7.02 & 14.38 & 13.54 \\
& Max & 87.59 & 105.94 & 98.58 \\
& Min & 00.22 & 17.96 & 07.13 \\ \hline

\multirow{4}{*}{\textbf{Speed (m/s)}} 
& Mean & 1.33 & 1.58 & 1.51 \\
& Std & 1.53 & 1.74 & 1.66 \\
& Max & 5.0 & 5.08 & 5.07 \\
& Min & 0.002 & 0.001 & 0.002 \\ \hline

\end{tabular}
\label{tab:race_statistics}
\vspace{-0.5cm}
\end{table}

In the DT, the actual UAV is emulated using a Software-In-The-Loop (SITL) vehicle emulator. Commands to move the UAV were written in the experiment code within E-VM, but they pass through a Control VM (C-VM) before reaching the UAV. The C-VM contains a safety filter that blocks commands not allowed by the AERPAW platform, such as commands violating the AERPAW geofence.
A representative diagram for real world testbed and DT is shown in Fig.~\ref{fig:afar_system}.
Each team designs a unique algorithm that follows different trajectories as their UAV searches for the hidden rover. During the flights, AERPAW records detailed telemetry, including timestamped 3D positions of the UAVs along with orientation (roll, pitch, and yaw angles). These records were included in the final dataset, offering rich information for post-analysis and learning.
\vspace{-0.3cm}
\section*{VALIDATION AND QUALITY} 

    % \textcolor{blue}{Validation and Quality must provide details that support the technical quality and/or accuracy of the data collected. For example, the error rate or accuracy of any hardware sensors used for collection. Authors should provide figures and tables to support this.}

    % \textcolor{red}{For this section, Dr. Ozgur Dr. Mihail and Anil input would be valuable in briefly commenting on the accuracy of the sensors and the collected measurements, including the accuracy of the UAV’s position (GPS location). Additionally, a brief comment on the accuracy of other equipment, such as the sensor used to collect UAV roll, yaw, and pitch information, as well as any aspects related to the antenna, RF hardware, or timestamp accuracy, would be helpful, given Dr. Ozgur’s expertise. If there are any figures that could be included to illustrate the accuracy aspects, please let me know, and I will take care of that. 
    % I noticed the time stamps are coming for some GPS time. Just writing briefly about it would be great. }

The AFAR dataset exhibits high completeness and consistency, with no missing entries or out-of-range values across its features. All recorded parameters remain within expected operational bounds. Table \ref{tab:race_statistics} presents the statistical characteristics of RSS, RSQ, and UAV speed for Team 288 across the three deployment locations. These results are shown as a representative example; for the other teams, the mean and standard deviation values differ due to variations in their designed UAV trajectories as well as additional factors such as multipath reflections and noise. Specifically, both RSS and RSQ exhibit a wide dynamic range, which arises not only from changes in the UAV’s distance to the RF source but also from propagation effects such as multipath fading, environmental shadowing, noise, and UAV body or orientation-induced blockage. The UAV speed during the flights ranges from approximately 0.001 m/s to 5 m/s. This variation is attributed to the waypoint-based trajectory design adopted by the teams. As the UAV reaches a waypoint, it momentarily decelerates or halts, resulting in near-zero speed readings. Subsequently, as it transitions to the next waypoint, the UAV accelerates, producing higher speed values. All other features (further detailed in Section IV) are similarly well-bounded and free from anomalies.

\begin{table}[!t] %H
\vspace{-0.6cm}
\caption{Statistics of received samples.}
\centering
\begin{tabular}{|l|l|c|c|c|}
\hline
\textbf{Feature} & \textbf{Stat.} & \textbf{Loc-1} & \textbf{Loc-2} & \textbf{Loc-3} \\ \hline

\multirow{4}{*}{\textbf{Team 288}} 
& Flight Time (sec) & 711 & 727 & 689 \\
& Expected \# of Samples & 21330 & 21810 & 20670 \\
& Total \# of Samples & 21714 & 22217 & 21056 \\ \hline

\multirow{4}{*}{\textbf{Team 300}} 
& Flight Time (sec) & 699 & 710 & 741 \\
& Expected \# of Samples & 20970 & 21300 & 22230 \\
& Total \# of Samples & 19147 & 19561 & 20388 \\ \hline

\multirow{4}{*}{\textbf{Team 301}} 
& Flight Time (sec) & 598 & 676 & 649 \\
& Expected \# of Samples & 17940 & 20280 & 19470 \\
& Total \# of Samples & 17630 & 18181 & 17473 \\ \hline

\multirow{4}{*}{\textbf{Team 309}} 
& Flight Time (sec) & - & 991 & 849 \\
& Expected \# of Samples & - & 29730 & 25470 \\
& Total \# of Samples & - & 29379 & 25163 \\ \hline

\multirow{4}{*}{\textbf{Team 328}} 
& Flight Time (sec) & 854 & 706 & 725 \\
& Expected \# of Samples & 25620 & 21180 & 21750 \\
& Total \# of Samples & 25909 & 21394 & 21956 \\ \hline

\end{tabular}
\label{tab:Samples}
\vspace{-0.4cm}
\end{table}

Throughout all experiments, RSS and RSQ measurements were consistently sampled at a fixed rate of 30 samples per second. Table \ref{tab:Samples} provides statistics on the sampling intervals and completeness of the collected data. It reports the total flight duration for each team and location, the expected number of samples based on the nominal sampling rate, and the actual number of samples recorded. In some instances, the number of collected samples slightly exceeds the expected count due to the equipment operating at a marginally higher sampling rate. Conversely, minor deficits in a few cases are observed but remain negligible, demonstrating the precision, reliability, and completeness of the data collection process. These variations do not affect time stamp accuracy, as they are system-clock-generated and aligned with actual sampling instances. For Team 309 at Location 1, the physical testbed dataset is unavailable due to accidental deletion during file transfer, although the localization estimate remains available. Locations 2 and 3 are intact, and the DT data for all three locations is fully available. Thus, only one physical dataset is missing, while overall coverage and reusability for benchmarking remain unaffected.

For Team 309 at Location 1, the dataset is unavailable due to accidental deletion during file transfer. While the localization estimate is still available, the raw data was lost due to a handling error, not related to any issue in data collection or recording.

\begin{figure}
\vspace{-0.3cm}
\includegraphics[width=0.7\linewidth]{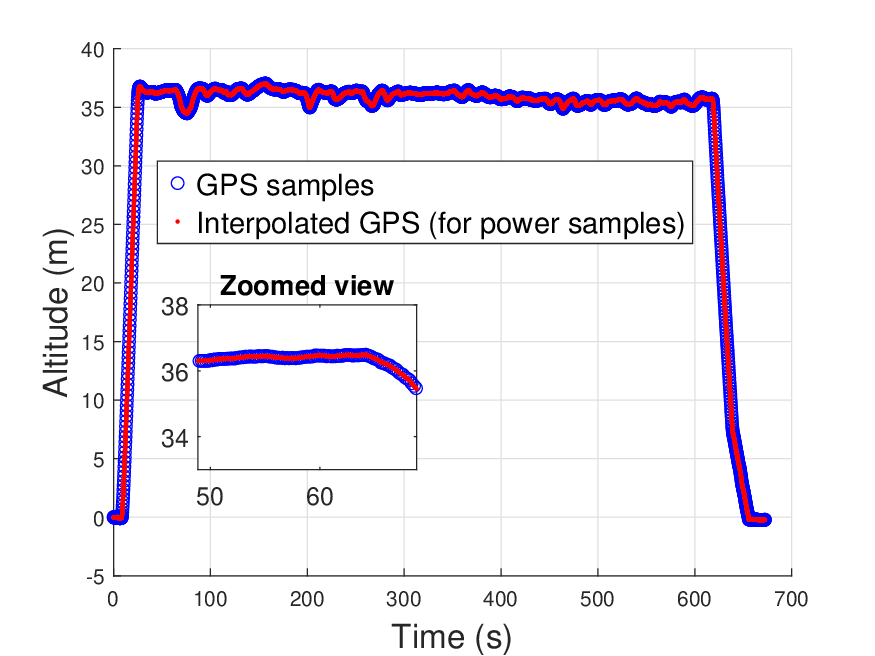}
	\centering
% \vspace{-0.1cm}
	\caption{Interpolation of sparse GPS altitude measurements onto RSS measurement timestamps. }
	\label{Interp}
\vspace{-0.9cm}
\end{figure}

The UAV state features, including longitude, latitude, altitude, speed, and absolute heading, were sampled at a lower rate of approximately 5 Hz through the flight computer. Positioning and timing are provided by an onboard real-time kinematic (RTK)-enabled global navigation satellite system (GNSS) system, with centimeter-level accuracy ensured by RTK corrections from AERPAW’s base station at Lake Wheeler. In contrast, the RSS and RSQ measurements were acquired at a higher rate of 30 Hz. This sampling disparity introduces the challenge of associating each RSS and RSQ sample with the corresponding spatial and kinematic information from the UAV. To address this, linear interpolation was employed to estimate the UAV state feature at the higher sampling rate of the RSS and RSQ measurements. For example, for each RSS sample timestamp, the corresponding UAV altitude was computed by interpolating between the available GPS altitude samples. This interpolation process was applied similarly to other UAV state features, ensuring temporal alignment across all recorded features. Fig.~\ref{Interp} illustrates the result of the interpolation process, where the original sparse GPS altitude samples (in blue) are complemented by interpolated values (in red). The interpolated points closely follow the trend of the original GPS samples, ensuring consistency of the UAV trajectory and reliable alignment of RSS/RSQ with state features.

% The original GPS altitude samples (blue circles) are linearly interpolated to obtain dense altitude estimates (red dots) aligned with the higher-rate RSS measurements.

\begin{figure}
% \vspace{-0.5cm}
\vspace{-0.3cm}
\includegraphics[width=0.7\linewidth]{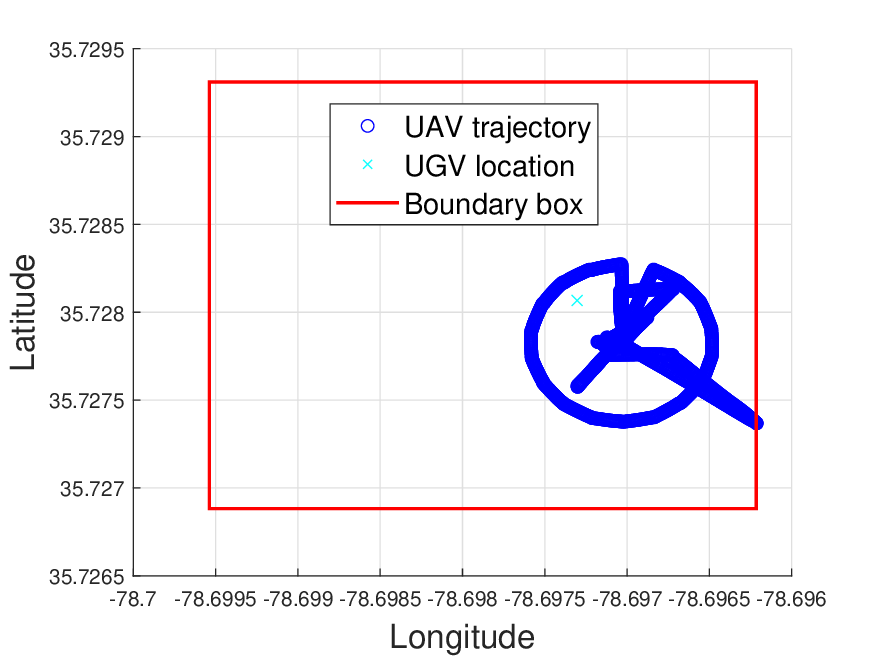}
	\centering
% \vspace{-0.1cm}
	\caption{UAV trajectory of Team-288 for Location-1, illustrating the UAV path within the predefined bounding box and the position of the RF source.}
	\label{GPS}
\vspace{-0.7cm}
\end{figure}

The GPS-derived coordinates of the UAV, including longitude, latitude, and altitude, are observed to be within the predefined flight boundary, as expected. Fig.~\ref{GPS} illustrates the UAV trajectory for Team-288 for the UGV located at Location-1. This visualization also highlights the UAV's maneuvering around the RF source, validating the integrity of the positional data. 

% \begin{table}[ht]
% \caption{Statistics From the Race Data, Regarding Speed (Both in Kilometer per Hour and in Knots) and Heading (in Degrees)}
% \centering
% \begin{tabular}{|l|l|c|c|c|}
% \hline
% \multicolumn{2}{|c|}{\textbf{Feature}} & \multicolumn{3}{c|}{\textbf{Race}} \\ \hline
%  & \textbf{Stat.} & \textbf{Speed} & \textbf{Avoidance} & \textbf{Endurance} \\ \hline

% \multirow{4}{*}{\textbf{Speed (kph)}} 
% & Mean & 3.18 & 2.22 & 3.70 \\
% & Std & 2.80 & 2.09 & 2.32 \\
% & Min & 0.02 & 0.02 & 0.02 \\
% & Max & 14.59 & 8.91 & 13.79 \\ \hline

% \multirow{4}{*}{\textbf{Speed (knots)}} 
% & Mean & 1.72 & 1.20 & 2.00 \\
% & Std & 1.51 & 1.13 & 1.25 \\
% & Min & 0.01 & 0.01 & 0.01 \\
% & Max & 7.88 & 4.81 & 7.45 \\ \hline

% \multirow{4}{*}{\textbf{Heading (degrees)}} 
% & Mean & 192.80 & 164.71 & 169.89 \\
% & Std & 97.54 & 103.43 & 104.29 \\
% & Min & 0.23 & 0.24 & 0.04 \\
% & Max & 359.70 & 359.22 & 359.84 \\ \hline

% \end{tabular}
% \label{tab:race_statistics}
% \end{table}

\section*{RECORDS AND STORAGE} 

% \textcolor{blue}{Generic Guidelines for this section:  Records and Storage must provide details on how the data files are structured and how the data is stored. A table that lists all files with summary details is recommended. For example, for CSV files, a description of the columns and rows of each file is needed. Additionally, details on how files relate to each other are needed, especially if there is some hierarchical structure. For example, file A is a summation of files B and C plus some error terms. The format of each data file should be sufficiently described.}

\begin{table}[!b]
\vspace{-0.6cm}
\centering
\caption{Description of primary data files in each location-specific folder.}
\label{table1}
\setlength{\tabcolsep}{3pt}
\renewcommand{\arraystretch}{1}  % Increases row spacing for better readability
\begin{tabular}{|p{53pt}|p{185pt}|}
 \hline
\textbf{File} & \textbf{Description} \\
% \hline
\hline
$log.csv$ & Contains time-synchronized UAV positional and state information, including longitude, latitude, altitude, and speed. The file includes 12 fields, each representing a critical flight parameter for detailed spatial and flight behavior analysis. \\
\hline
$power\_log.txt$ & Records time-synchronized RSS values. Contains three columns: the first for timestamps and the third for RSS.
 \\
\hline
$quality\_log.txt$ & Captures RSQ values with timestamps. Structured similarly to the power log. \\
\hline
$angles.mat$ & Stores the UAV’s orientation—roll, pitch, and yaw angles.\\
\hline
\end{tabular}
\vspace{-0.4cm}
\end{table}

While the primary focus of this paper is on the analysis and exploration of the real-world dataset, both the DT and real-world datasets share a unified structure. The dataset is organized into directories based on each team’s experiment ID (T-XXX). Within each team folder, there are two subfolders: ‘development’, containing the DT-collected data, and ‘testbed ’, containing the real-world data. Each of these subfolders is further divided into three location-specific directories—loc-1, loc-2, and loc-3, corresponding to different UGV placements (e.g., 288/loc-1, 288/loc-2, 288/loc-3).  For the real-world dataset located in the ‘testbed' folder, each location-specific directory contains four primary files: ‘$log.csv$’, ‘$power\_log.txt$’, ‘$quality\_log.txt$`, and ‘$angles.mat$’, as summarized in Table~\ref{table1}.

%---------- Imp ------------------
Table~\ref{tablelogs} provides a detailed description of the fields contained in the $log.csv$ file, which records the UAV’s navigation and positional data throughout each flight. These fields include highly time-synchronized GPS-derived coordinates (latitude, longitude, and altitude), speed metrics, and satellite information. The combination of temporal and spatial parameters allows for the accurate reconstruction of the UAV's flight trajectory and enables synchronization with RF measurements such as RSS and RSQ. Table~\ref{Powerlog} outlines the structure of the power\_log.txt file, which contains RSS measurements captured by the UAV during flight. The quality\_log.txt file follows an identical format, with the third column containing RSQ values instead of RSS. Together, these files offer a comprehensive dataset for characterizing and analyzing A2G wireless channel performance.

\begin{table}[!t]
\vspace{-0.5cm}
\centering
\caption{Description of key fields in $log.csv$, capturing UAV state features. }
\label{tablelogs}
\setlength{\tabcolsep}{3pt}
\renewcommand{\arraystretch}{1}  % Increases row spacing for better readability
\begin{tabular}{|p{50pt}|p{175pt}|}
 \hline
\textbf{File} & \textbf{Description} \\
% \hline
\hline
timestamp & Unix timestamp with microsecond precision.\\
\hline
TimeUS & Time since system startup (in seconds). \\
\hline
Status & Indicate the type and precision of the GPS.\\
\hline
GMS & Milliseconds since the start of the current GPS week.\\
\hline
GWK & GPS week count.  \\
\hline
NSats & Number of visible satellites.\\
\hline
HDop & Accuracy of horizontal GPS position.\\
\hline
Lat & Latitude of UAV (in degrees). \\
\hline
Lon & Longitude of UAV (in degrees). \\
\hline
Alt & Altitude of UAV (in meters). \\
\hline
Spd & UAV ground speed (in m/s). \\
\hline
GCrs & Ground course or heading (in degrees). \\
\hline
VZ & Vertical speed (ascent/descent rate).\\
\hline
\end{tabular}
\vspace{-0.4cm}
\end{table}

\begin{table}[!t]
\vspace{-0.3cm}
\centering
\caption{Description of fields in \texttt{power\_log.txt} file.}
\label{Powerlog}
\setlength{\tabcolsep}{3pt}
\renewcommand{\arraystretch}{1}  % Increases row spacing for better readability
\begin{tabular}{|p{50pt}|p{180pt}|}
 \hline
\textbf{File} & \textbf{Description} \\
% \hline
\hline
First Column & Timestamp corresponding to each RSS measurement, recorded with microsecond precision.\\
\hline
Second Column & Sample index or counter value; not required for signal analysis and may be disregarded.\\
\hline
Third Column & RSS measured by the UAV receiver, used for assessing link strength and propagation characteristics. \\
\hline

\end{tabular}
\vspace{-0.3cm}
\end{table}

Table~\ref{anglelogs} summarizes the UAV orientation parameters available in the angles.mat file. These values enable precise attitude estimation, which is essential for modeling the shadowing effects and understanding the impact of UAV dynamics on signal propagation and localization performance. 

\begin{table}[!t]
\vspace{-0.2cm}
\centering
\caption{Description of variables in \texttt{angles.mat} file. }
\label{anglelogs}
\setlength{\tabcolsep}{3pt}
\renewcommand{\arraystretch}{1}  % Increases row spacing for better readability
\begin{tabular}{|p{25pt}|p{210pt}|}
 \hline
\textbf{File} & \textbf{Description} \\
% \hline
\hline
mRoll & Rotation of the UAV about its longitudinal (front-to-back) axis, expressed in degrees.\\
\hline
mPitch & Rotation of the UAV about its lateral (side-to-side) axis, in degrees. Captures the upward or downward tilt.\\
\hline
mYaw & Rotation of the UAV about its vertical axis, in degrees. Describes the heading or directional orientation of the UAV.  \\
\hline

\end{tabular}
\vspace{-0.3cm}
\end{table}

% Each team has its own UAV trajectory, where the UAV is flying at different heught to collect signals for the UGV, Fig. \ref{fig:FullPage} first columns shows the height at which the UAV is flying for all five team for loc2.

%----------------Imp---------
\begin{figure*}
    \centering
    % First row
    \vspace{-7mm}
    \begin{subfigure}{0.3\textwidth}
        \centering
        \includegraphics[width=\linewidth]{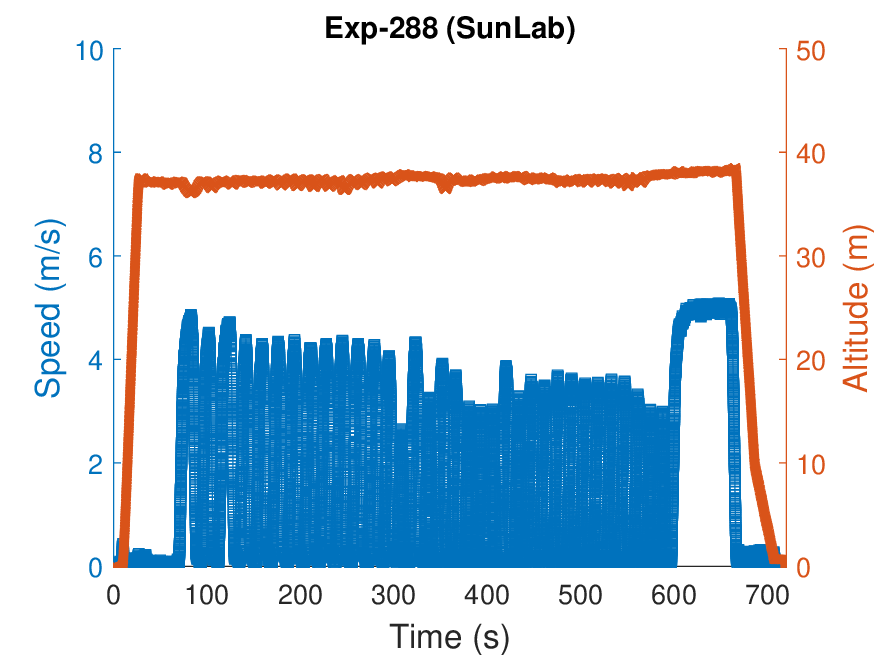}
        \caption{UAV altitude and speed over time.}
        \label{fig:Label1}
    \end{subfigure}
    \hspace{1.5mm}
    \begin{subfigure}{0.3\textwidth}
        \centering
        \includegraphics[width=\linewidth]{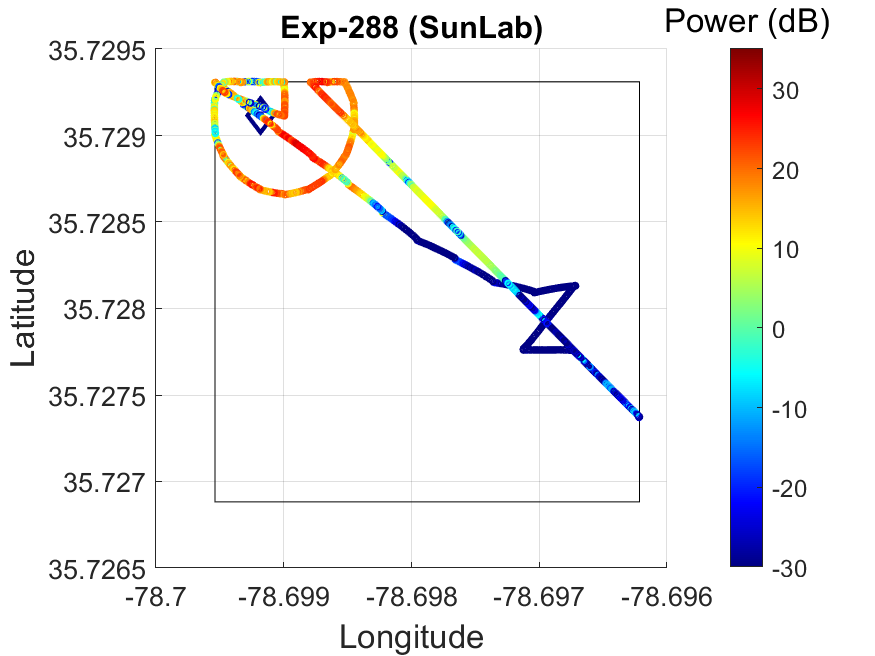}
        \caption{UAV trajectory and received RSS.}
        \label{fig:Label2}
    \end{subfigure}
    \hspace{1.5mm}
    \begin{subfigure}{0.3\textwidth}
        \centering
        \includegraphics[width=\linewidth]{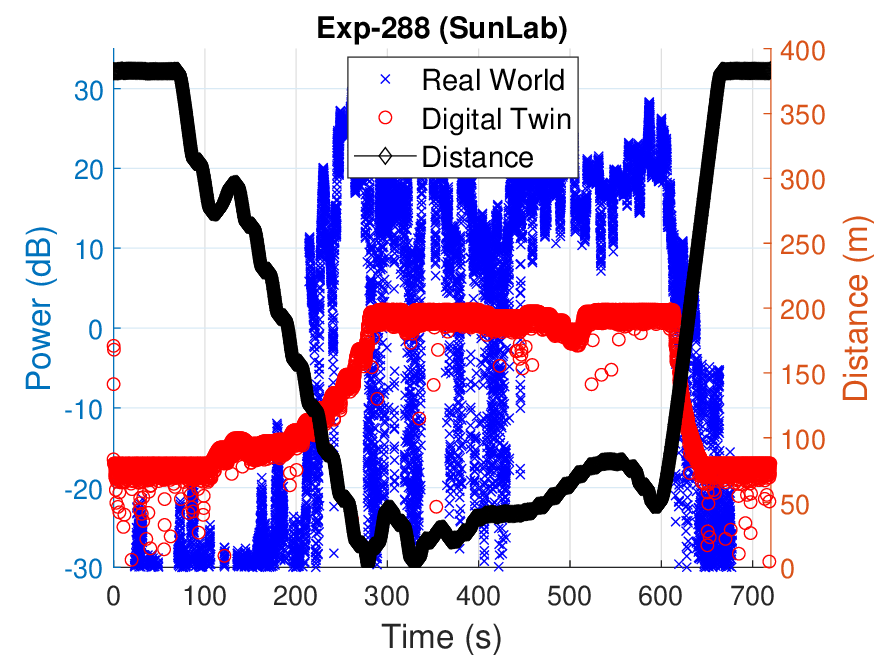}
        \caption{DT vs. real world RSS and distance}
        \label{fig:Label3}
    \end{subfigure}

    \vspace{-0.5mm} % Adjust vertical spacing

    % Second row
    \begin{subfigure}{0.3\textwidth}
        \centering
        \includegraphics[width=\linewidth]{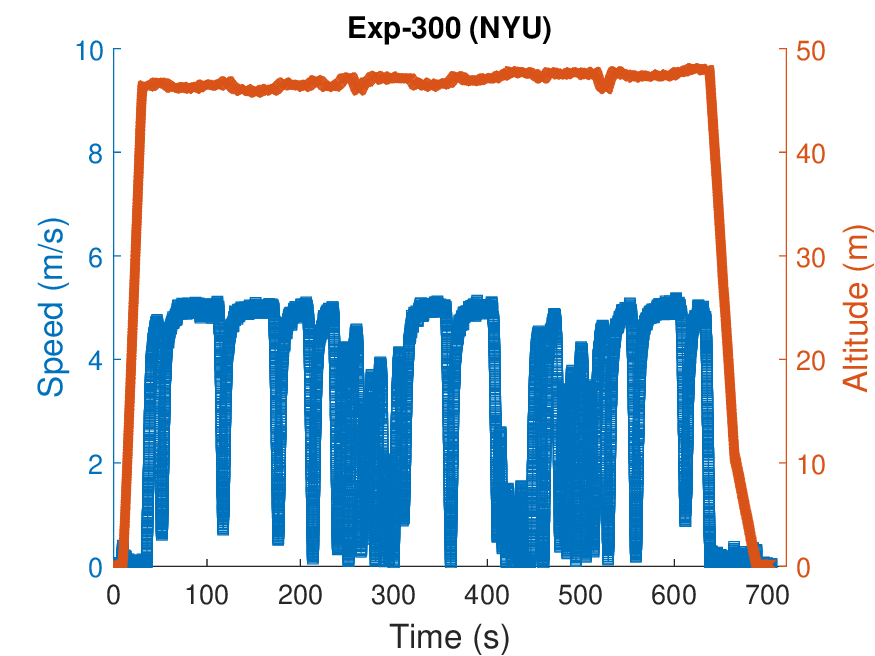}
        \caption{UAV altitude and speed over time.}
        \label{fig:Label4}
    \end{subfigure}
    \hspace{1.5mm}
    \begin{subfigure}{0.3\textwidth}
        \centering
        \includegraphics[width=\linewidth]{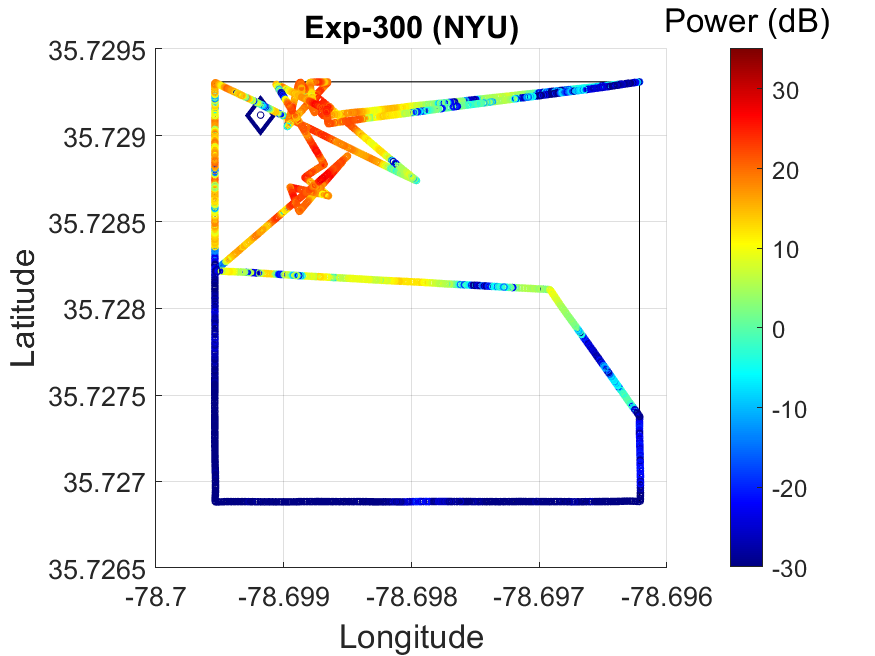}
        \caption{UAV trajectory and  RSS.}
        \label{fig:Label5}
    \end{subfigure}
    \hspace{1.5mm}
    \begin{subfigure}{0.3\textwidth}
        \centering
        \includegraphics[width=\linewidth]{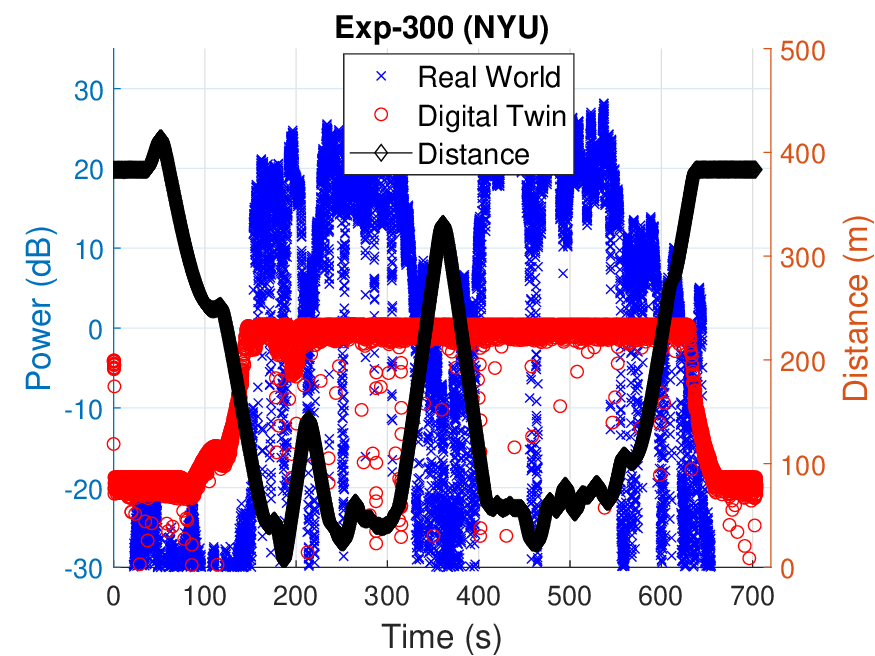}
        \caption{DT vs. real world RSS and distance}
        \label{fig:Label6}
    \end{subfigure}

    \vspace{-0.5mm} % Adjust vertical spacing

    % Third row
    \begin{subfigure}{0.3\textwidth}
        \centering
        \includegraphics[width=\linewidth]{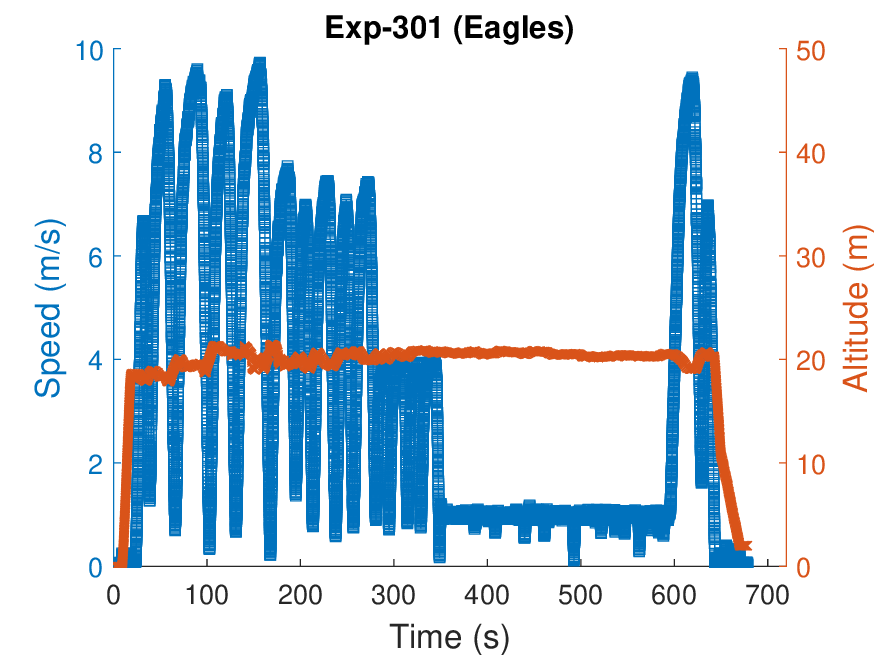}
        \caption{UAV altitude and speed over time.}
        \label{fig:Label7}
    \end{subfigure}
    \hspace{1.5mm}
    \begin{subfigure}{0.3\textwidth}
        \centering
        \includegraphics[width=\linewidth]{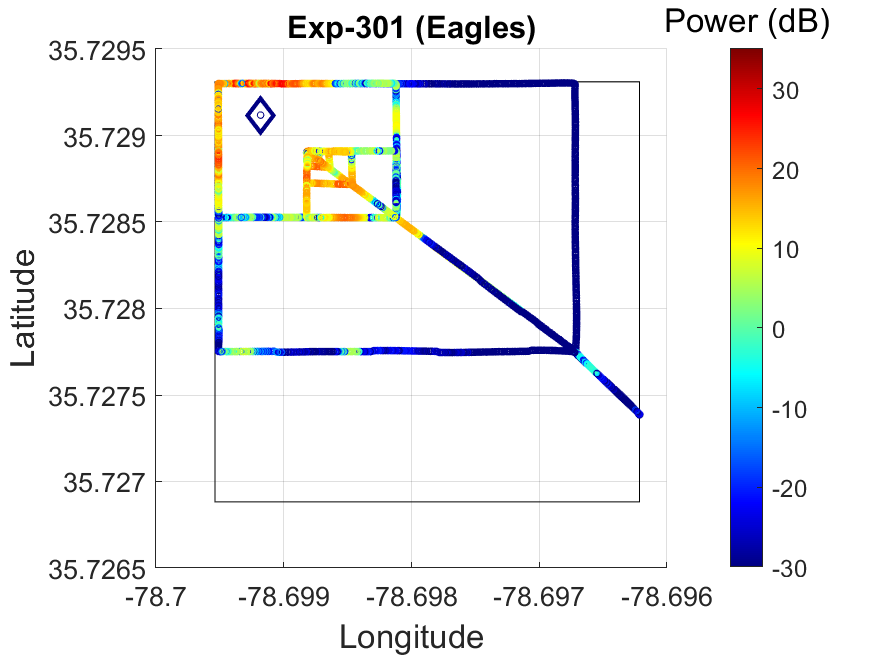}
        \caption{UAV trajectory and RSS.}
        \label{fig:Label8}
    \end{subfigure}
    \hspace{1.5mm}
    \begin{subfigure}{0.3\textwidth}
        \centering
        \includegraphics[width=\linewidth]{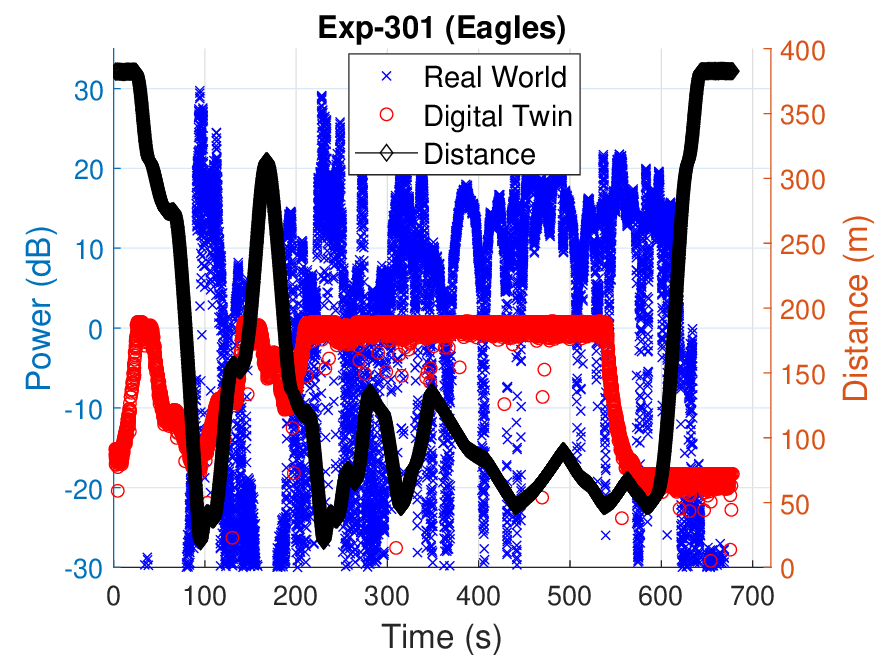}
        \caption{DT vs. real world RSS and distance}
        \label{fig:Label9}
    \end{subfigure}

    \vspace{-0.5mm} % Adjust vertical spacing

    % Fourth row
    \begin{subfigure}{0.3\textwidth}
        \centering
        \includegraphics[width=\linewidth]{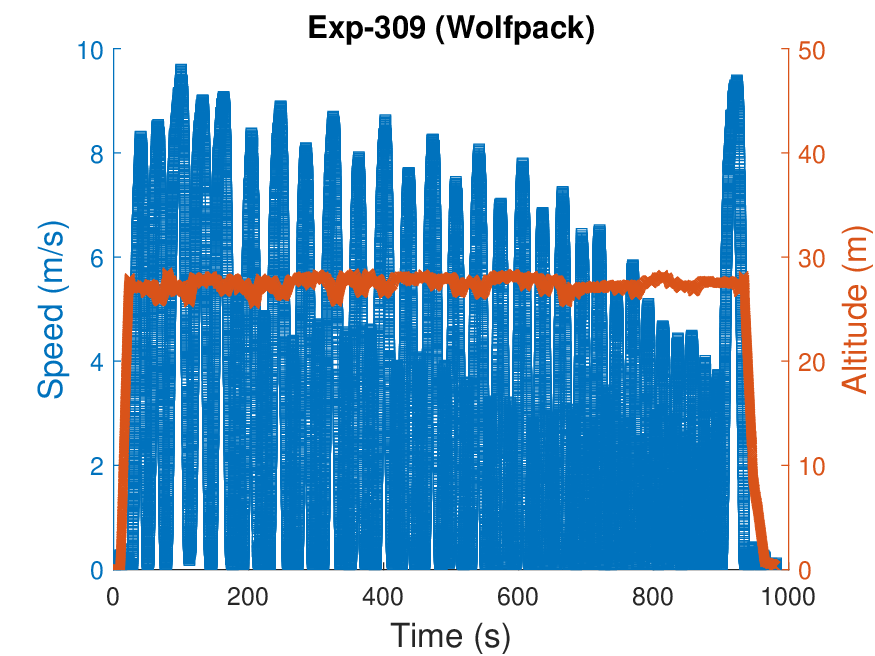}
        \caption{UAV altitude and speed over time.}
        \label{fig:Label10}
    \end{subfigure}
    \hspace{1.5mm}
    \begin{subfigure}{0.3\textwidth}
        \centering
        \includegraphics[width=\linewidth]{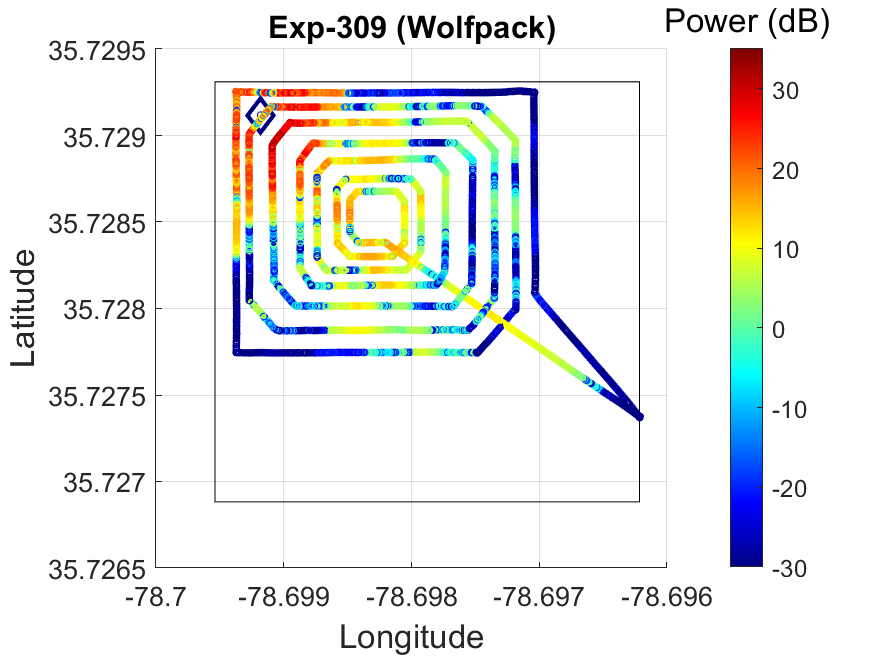}
        \caption{UAV trajectory and RSS.}
        \label{fig:Label11}
    \end{subfigure}
    \hspace{1.5mm}
    \begin{subfigure}{0.3\textwidth}
        \centering
        \includegraphics[width=\linewidth]{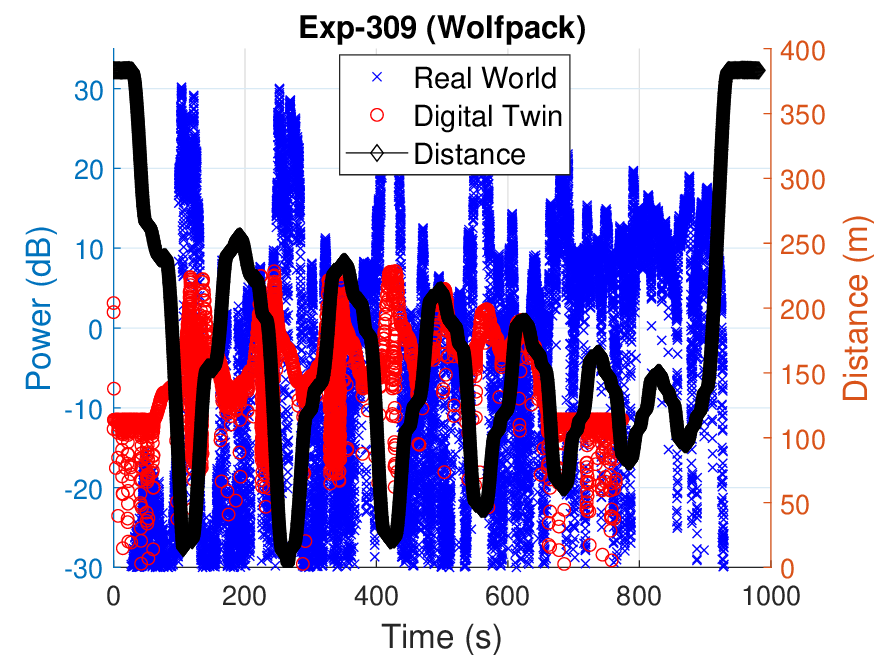}
        \caption{DT vs. real world RSS and distance.}
        \label{fig:Label12}
    \end{subfigure}

    \vspace{-0.5mm} % Adjust vertical spacing

    % Fifth row
    \begin{subfigure}{0.3\textwidth}
        \centering
        \includegraphics[width=\linewidth]{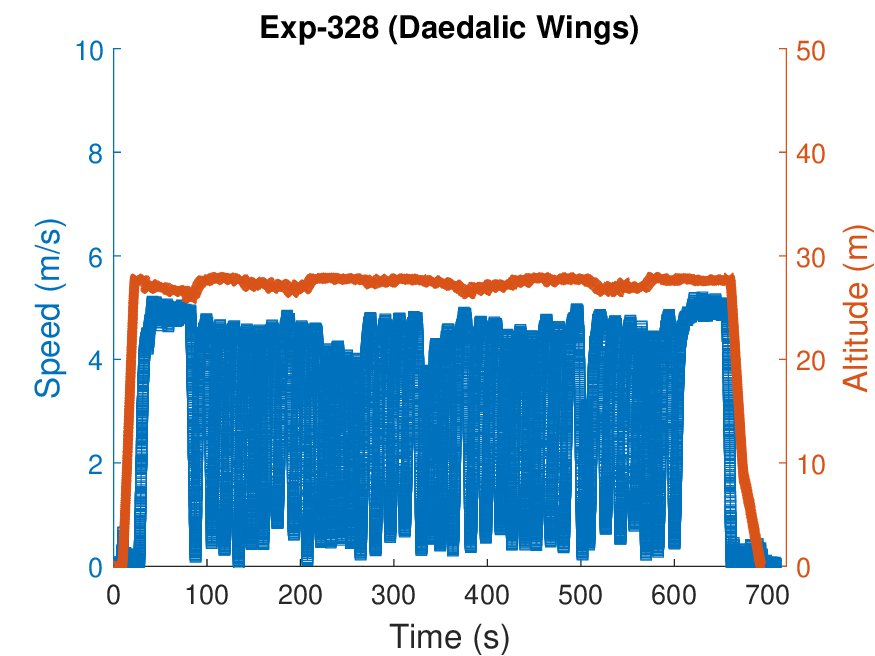}
        \caption{UAV altitude and speed over time.}
        \label{fig:Label13}
    \end{subfigure}
    \hspace{1.5mm}
    \begin{subfigure}{0.3\textwidth}
        \centering
        \includegraphics[width=\linewidth]{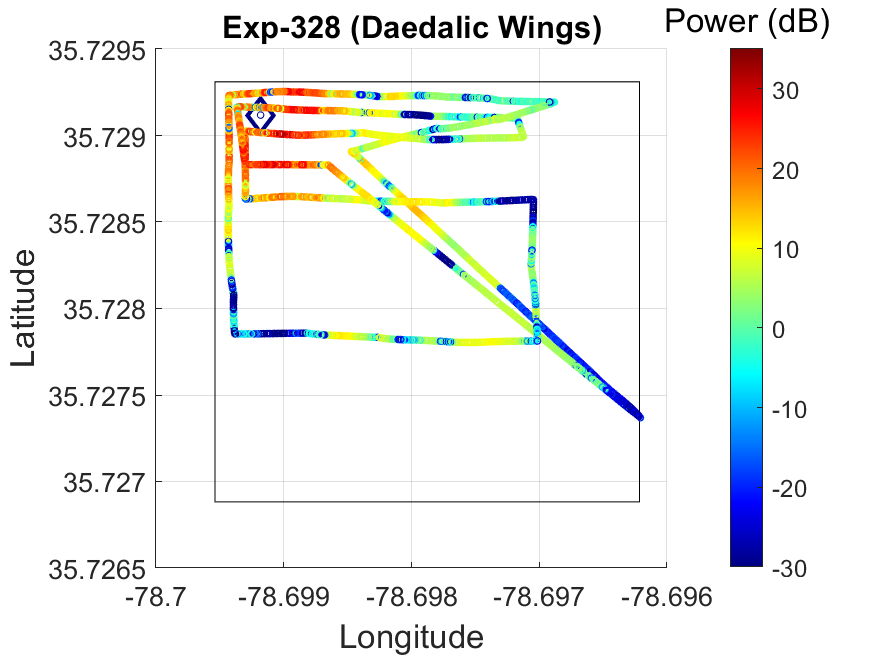}
        \caption{UAV trajectory and RSS.}
        \label{fig:Label14}
    \end{subfigure}
    \hspace{1.5mm}
    \begin{subfigure}{0.3\textwidth}
        \centering
        \includegraphics[width=\linewidth]
        {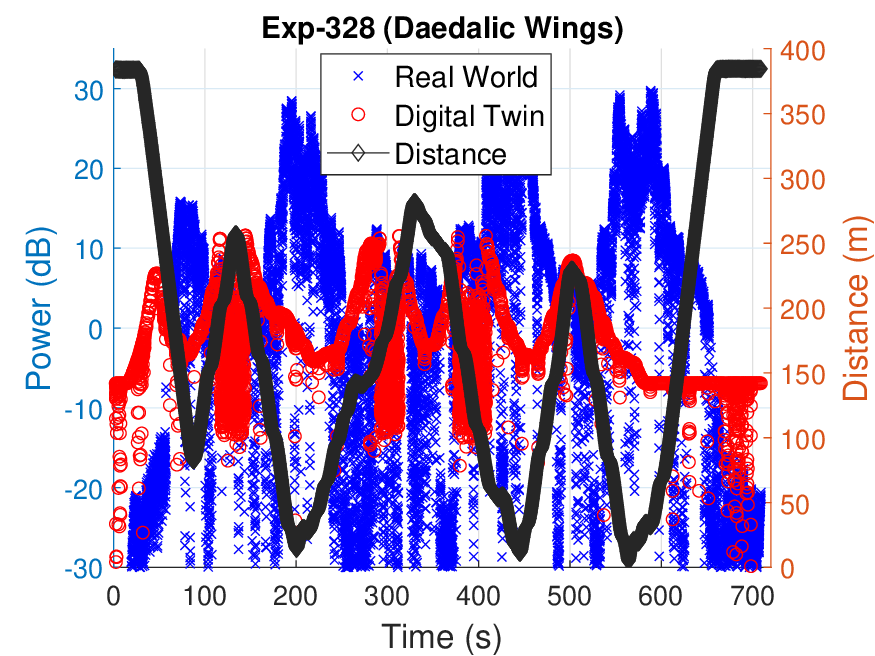}
        \caption{DT vs. real world RSS and distance.}
        \label{fig:Label15}
    \end{subfigure}

    \vspace{-2mm} % Adjust spacing if needed
    \caption{Each row in the figure corresponds to data collected from different experiments, starting from Exp288–Exp328.}
    \label{fig:FullPage}
    \vspace{-5mm} % Adjust spacing if needed
\end{figure*}
%------IMp------------------

\begin{figure*} % Use figure* to span both columns
\vspace{-0.4cm}
    \centering
    % First subfigure
    % \captionsetup[subfigure]{skip=0pt}  % Removes extra vertical spacing for subfigures
    \begin{subfigure}[b]{0.22\textwidth}
        \centering
        \includegraphics[width=\textwidth]{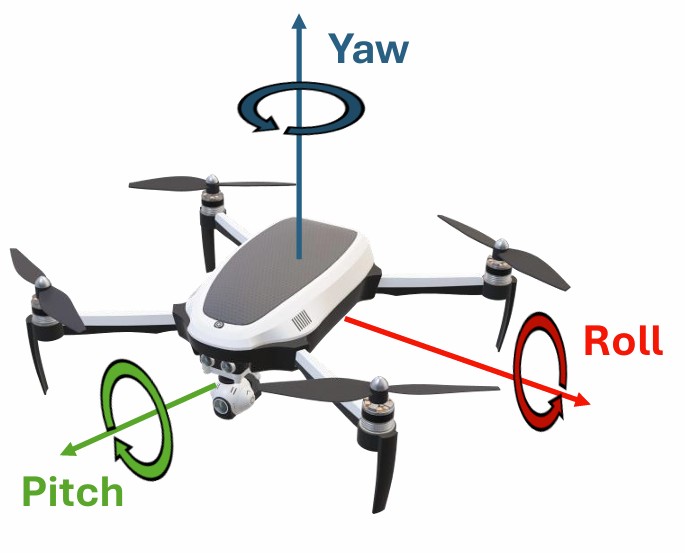}
        \caption{UAV's roll, pitch, and yaw.}
        \label{fig:CA4}
    \end{subfigure}
    \hspace{1.5mm}
    \begin{subfigure}[b]{0.22\textwidth}
        \centering
        \includegraphics[width=\textwidth]{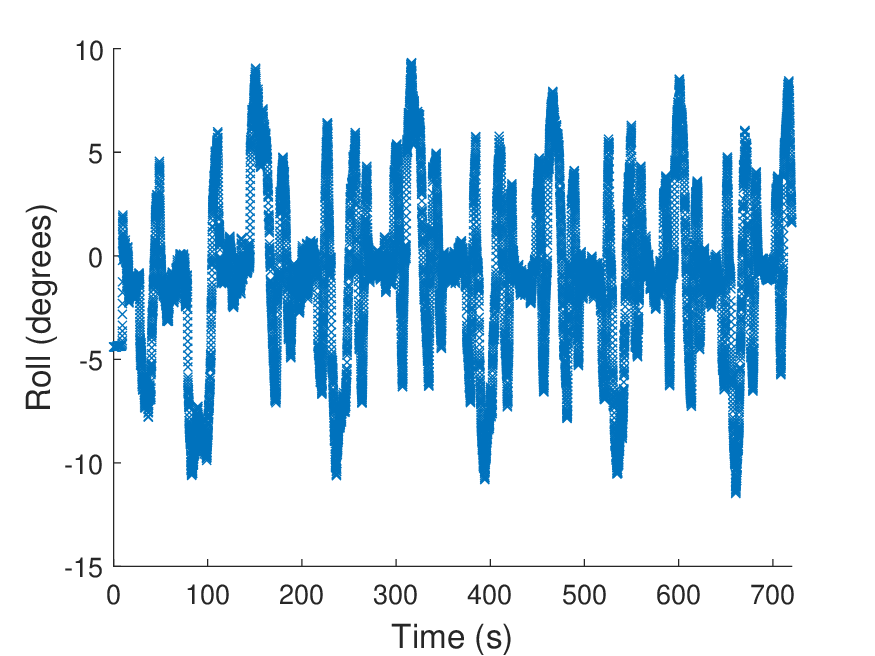}
        \caption{Roll}
        \label{fig:CA1}
    \end{subfigure}
    \hspace{1.5mm}
    % Second subfigure
    \begin{subfigure}[b]{0.22\textwidth}
        \centering
        \includegraphics[width=\textwidth]{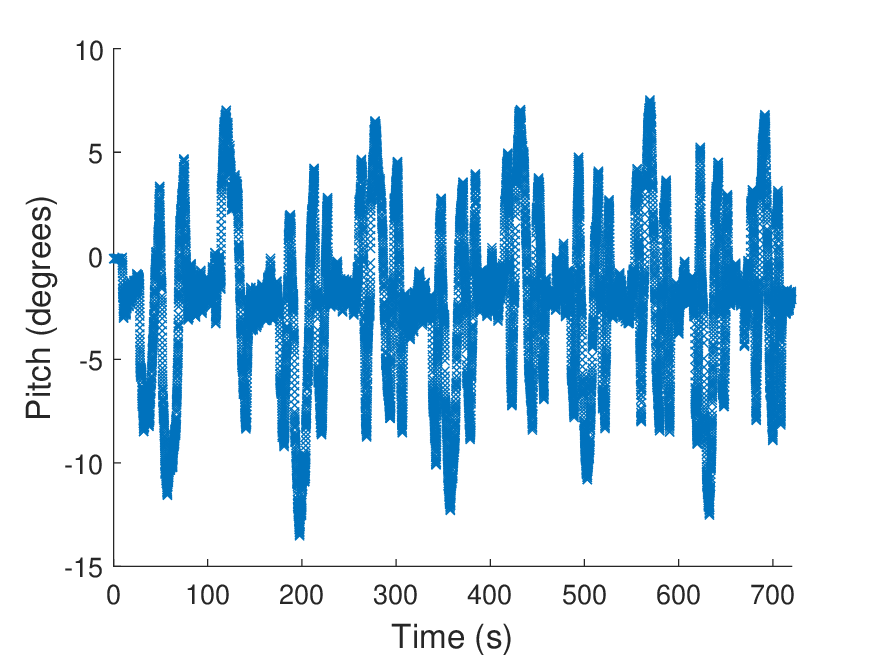}
        \caption{Pitch}
        \label{fig:CA2}
    \end{subfigure}
    \hspace{1.5mm}
    % Third subfigure
    \begin{subfigure}[b]{0.22\textwidth}
        \centering
        \includegraphics[width=\textwidth]{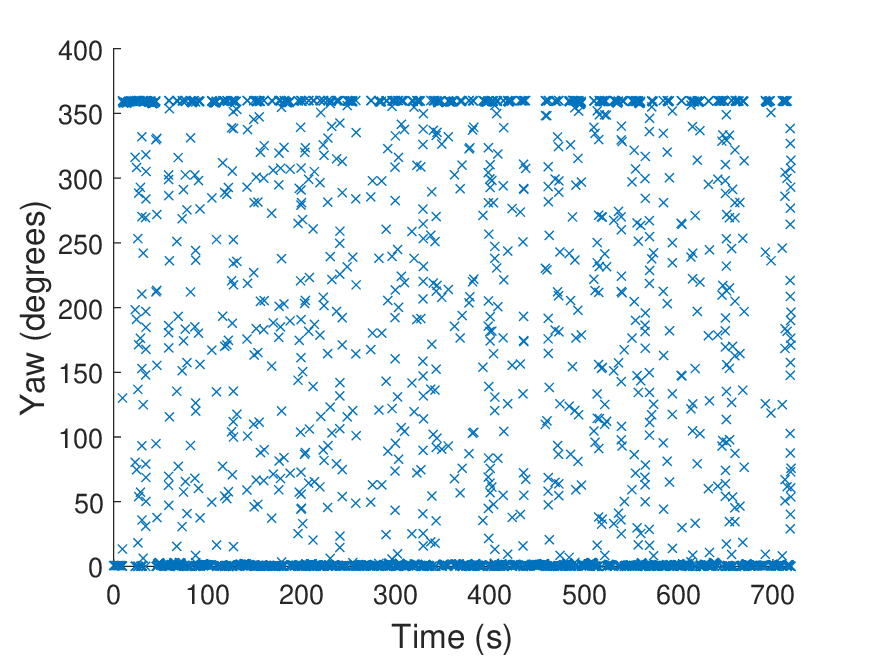}
        \caption{Yaw}
        \label{fig:CA3}
    \end{subfigure}
    % Main figure caption
    \vspace{-3mm}
    \caption{Time-series plots of the UAV's roll, pitch, and yaw angles for Team 288, when the UGV is at Location 1.}
    \label{fig:Orientation}
    \vspace{-7mm} % Adjust this value if needed
\end{figure*}

\section*{INSIGHTS AND NOTES}

The AFAR dataset is a highly versatile and comprehensive resource for advancing UAV-assisted wireless communication research. An illustrative example of the AFAR dataset’s richness and utility is presented in Fig.~\ref{fig:FullPage}, where multiple features—including UAV speed, altitude, trajectory, distance to the RF source, and RSS—are visualized for all participating teams at Loc-2. This figure offers detailed insights into the dataset's structure and demonstrates how it can be leveraged to explore the relation between key variables.
The first column of the figure displays UAV altitude and speed as functions of time. The second column presents a heatmap of RSS values overlaid on the UAV’s flight trajectory, illustrating spatial variations in RSS across the area of operation. The third column compares real-world and DT RSS measurements over time, alongside the temporal variation in distance between the UAV and the RF source.

From the first column in Fig.~\ref{fig:FullPage}, it can be observed that each team employed different altitudes for UAV flight during their respective trajectories. In all cases, the UAV ascends from its starting position at the beginning of the flight and descends near the end, indicating the takeoff and landing phases. Outside of these transitions, the UAV typically maintains a constant altitude. Altitude plays a significant role in RSS behavior—higher altitudes improve LoS conditions; however, greater altitude also increases the distance to the RF source, which can lead to signal attenuation. The AFAR dataset allows analysis of RSS variation with UAV altitude, providing real-world insights.

The same column also presents the UAV’s speed profile, which exhibits a consistent pattern of increasing and decreasing speed over time. The team's UAV exhibits a speed range from 0 to 10 m/s, reflecting diverse operational states and maneuvering behaviors throughout the flight.  This is attributed to the waypoint-based UAV trajectory.  UAV trajectories in AFAR are of two types: fixed waypoints and autonomous waypoints. In the fixed approach, teams predefine a set of waypoints in a plan file that the UAV follows sequentially. In contrast, the autonomous approach involves decision-making algorithms that dynamically determine the next waypoint based on the current state and measurements. These two strategies lead to distinct movement patterns and influence the UAV's interaction with the RF environment, as can be seen in the second column of Fig.~\ref{fig:FullPage}. The UAV accelerates while traveling toward a waypoint and decelerates upon arrival, where it may pause briefly before proceeding, causing a decrease in the speed, which is visible in Fig.~\ref{fig:FullPage}. This behavior is consistent across teams and trajectories. UAV speed also has an important impact on signal measurement. Using the AFAR dataset, researchers can investigate the relationship between UAV speed, RSS fluctuations, and overall system performance, enabling more informed and adaptive algorithm development.

The second column of Fig.~\ref{fig:FullPage} illustrates the UAV trajectories for all teams, with the first three teams employing autonomous and the last two using fixed waypoint trajectories. A heatmap of RSS is overlaid along the UAV paths, providing a spatial visualization of signal variations. The diamond marker in each plot denotes the position of the RF source. As expected, RSS is highest when the UAV is near the RF source and decreases with distance. This representation offers insight into how different trajectory strategies impact signal behavior and localization performance. The ability to compare autonomous versus fixed trajectory planning within the same environmental conditions allows researchers to assess which approach is more effective for RF source tracking and A2G channel modeling. Moreover, the heatmap visualization highlights both temporal and spatial variations in RSS, providing a rich foundation for studying trajectory-dependent signal propagation characteristics.

The third column of Fig.~\ref{fig:FullPage} presents the RSS and the UAV-to-RF source distance as functions of time. A clear distinction can be observed between the DT and real-world measurements. In the DT environment, RSS follows a relatively smooth, distance-dependent trend, whereas the real-world data exhibit significant fluctuations even at similar distances, highlighting the complexity and variability of RF propagation. These variations arise from probabilistic LoS conditions, multipath reflections, environmental shadowing, deep fades, and UAV body-induced blockage. While a general trend of stronger RSS at shorter distances is visible, the real-world data clearly demonstrate that distance alone cannot fully explain signal behavior, challenging the assumptions of conventional path loss models. The DT environment is based on a two-ray propagation model with one LoS and one ground-reflected NLoS path, combined with Gaussian noise. It replays the same UAV trajectories and software as the real testbed, with transmit power, transmitter/receiver gains, and trajectory parameters calibrated to match the physical experiments. However, this simplified model cannot capture additional reflections, hardware imperfections, or dynamic environmental factors observed in practice.

Quantitatively, for Team T-$328$, the overall mean RSS is identical in both domains ($–1.29$ dB), but the standard deviation (std) is much higher in the real-world data ($18.09$~dB) compared to the DT ($5.53$~dB). Similarly, at a distance of approximately $150$±$5$~m, the real-world RSS has a mean of $–5.92$~dB with a std of $15.17$~dB, while the DT reports $3.16$~dB with a std of only $5.59$~dB. These results highlight that DT captures the large-scale trend, while the real-world data introduces substantial variability due to fading and environmental influences. Therefore, the two domains should be seen as complementary resources: DT data provide scalable, controlled measurements suitable for pretraining and benchmarking, while real-world datasets enable fine-tuning and validation under realistic conditions. Leveraging both jointly is especially valuable for deep learning workflows, such as transfer learning and domain generalization.

Fig.~\ref{fig:Orientation} illustrates the UAV’s orientation—roll, pitch, and yaw over time, providing valuable insights into flight dynamics and body-induced signal shadowing. Since the receive antenna is mounted on the UAV, changes in orientation significantly affect the antenna pattern, leading to variations in received RSS. In \cite{masrur2025bridging}, these effects were analyzed using the AFAR dataset. The study highlighted that due to the vertical polarization of the receiving antenna, orientation changes cause polarization mismatch and alter antenna gain, impacting signal reception. In our setup, both the UAV receiving antenna and the UGV transmitting antenna are vertically polarized.

\begin{figure}
\vspace{-0.4cm}
\includegraphics[width=0.8\linewidth]{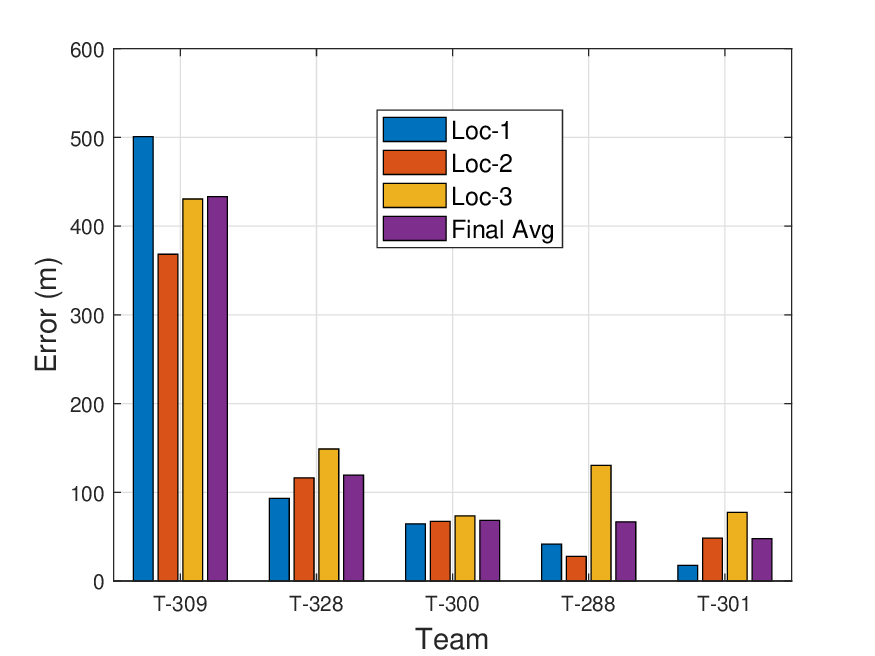}
	\centering
% \vspace{-0.1cm}
	\caption{Localization errors at each UGV location and the final average.}
	\label{localization}
\vspace{-0.5cm}
\end{figure}

The localization error for each team is shown in Fig.~\ref{localization}. Team T-301 achieved the highest accuracy, winning the competition with a final average localization error of 47.8 meters. Team T-288 secured second place with an average error of 66.6 meters, followed by Team T-300 in third place with an error of 68.3 meters. The participating teams employed diverse localization strategies. Team T-301, the winner, used a recursive perimeter-sweep algorithm with averaging to mitigate noise, achieving the best overall accuracy \cite{kudyba2024uav}. Team T-288 adopted an online Bayesian optimization framework with Gaussian Process (GP) surrogate modeling, leveraging outlier filtering and adaptive waypoint generation \cite{kudyba2024uav,10757589}. Team T-300 also relied on GP regression with Bayesian optimization, supplemented by a perimeter survey, but its strong simulation results did not generalize as well to the noisier real-world trials \cite{kudyba2024uav}. Team T-309 designed a fixed coverage trajectory combined with a particle filter, which proved sensitive to mismatches between assumed path loss models and real-world propagation. Team T-328 applied a least-squares method based on a path loss model using sampled RSS values, offering a simple but less robust approach \cite{10619824}. Localization error at Loc-3 is consistently higher across all teams compared to the other two locations. This is due to the RF source being positioned outside the UAV flight zone, preventing direct flyovers. These varied strategies demonstrate the complementary trade-offs between robustness, adaptability, and computational complexity, and provide useful insights for future research on UAV-based RF localization.

% Detailed descriptions of the algorithms used by these top-performing teams—including their localization strategies, trajectory planning approaches, and underlying motivations—are provided in \cite{kudyba2024uav,10757589}. Implementation details for Team T-328 can be found in \cite{}. Team T-309 adopted a fixed trajectory designed to sweep the entire search area within the allotted flight time, ensuring systematic coverage and avoiding unobserved regions. They employed a particle filter-based localization algorithm to estimate the RF source location. 

To address the change in antenna gains due to UAV orientation, the authors in \cite{masrur2025bridging} proposed an enhanced two-ray path loss model that incorporates antenna orientation and UAV body shadowing. Using this refined model, they trained a ResNet-based neural network for localization. The enhanced modeling approach significantly reduced the average localization error from 44 meters (achieved by the top-performing team in the AFAR Challenge) to 18.45 meters. This highlights the importance of accounting for orientation-induced antenna effects and encourages the development of more robust algorithms for advancing UAV-assisted wireless communication systems.

The dataset is subject to certain limitations. Environmental factors such as multipath reflections, foliage, and ground clutter can introduce variability in RSS and RSQ. While the data were collected in a rural environment, RSS and RSQ fluctuations may differ in urban deployments where dense clutter, frequent NLoS conditions, and strong reflections are more pronounced.

% The AFAR dataset offers a comprehensive and versatile foundation for advancing research in UAV-assisted wireless communication. It supports a wide range of applications, including A2G channel modeling, realistic path loss analysis, environmental factor characterization, and localization algorithm validation. The dataset enables researchers to quantify and reduce the gap between simulated and physical deployments, accelerating innovation in 5G/B5G connectivity, autonomous aerial systems, and UAV-based localization. Beyond traditional use cases, the dataset’s rich, time-synchronized measurements unlock opportunities in mobility-aware signal prediction, adaptive trajectory planning, and channel coherence analysis. 

Beyond traditional use cases, the AFAR dataset’s extensive, time-synchronized measurements offer significant opportunities for applying ML and DL techniques. Its scale and richness enable the development of data-driven models for mobility-aware signal prediction, adaptive trajectory planning, and channel coherence analysis. Researchers can train advanced ML/DL algorithms for tasks such as signal strength estimation, trajectory classification, anomaly detection, antenna pattern prediction, and deep fade forecasting—leveraging the dataset’s diversity across teams, trajectories, and environmental conditions. Moreover, the dataset allows for detailed analysis of how UAV trajectory impacts localization accuracy, enabling the design of more robust and efficient flight paths that can identify RF sources in reduced time. This makes AFAR a valuable testbed for learning-based approaches aimed at improving the performance and adaptability of UAV-assisted wireless communication systems in complex, real-world scenarios.

% by the teams. From my perspective, the data was not actively used by any team during the challenge; it was primarily collected as part of the process. I can explain the approach taken by each team, which mainly focused on data collection.

\section*{SOURCE CODE AND SCRIPTS} 

% \textcolor{red}{Source Code and Scripts must provide details on any public source code repositories that contain code used to collect, clean, or process this data. Details on any third-party software used should also be listed (including software version numbers). URLs and links to these repositories should be listed in the reference section. It is recommended that these repositories have an associated DOI to ensure their permanence: \url{https://github.com/some-user/some-repo/}.}

The data collection for this dataset was carried out using the AERPAW platform. Specifically, the GE2 experiment, as documented in AERPAW’s public repository \cite{aerpaw_ge2}, was used to configure UAV parameters and capture RF measurements during flight. Following data collection, the complete dataset was curated and publicly shared on Dryad:  
\url{https://doi.org/10.5061/dryad.18931zd4g}. 

The publicly available dataset folder contains a MATLAB script ($main.m$) designed to assist users with basic data parsing, cleaning, and post-processing tasks. These scripts allow users to extract core features such as timestamped RSS, RSQ, GPS coordinates, and UAV orientation (roll, pitch, yaw) from the raw log files, converting them into structured formats suitable for analysis.
Other datasets collected using the AERPAW platform are also publicly available at:
\url{https://aerpaw.org/experiments/datasets/}

\section*{ACKNOWLEDGEMENTS}
% The authors would like to thank Unmanned Experts, Galaxy Unmanned Systems, and AnyMile for being student award sponsors. 
The authors thank Unmanned Experts, Galaxy Unmanned Systems, and AnyMile for sponsoring student awards. The article authors have declared no conflicts of interest.

S. Masrur, Ö. Özdemir, and A. Gürses led the manuscript writing and organization. All other authors contributed to editing and review. Five independent teams participated in the AFAR Challenge. Contributors to algorithm development, UAV trajectory planning, and localization methods include S. Masrur, C. Dickerson, G. Reddy, S. Vargas Villar, C.-W. Wong; B. Chatterjee, S. Chaudhari, Z. Li, and Y. Liu; P. Kudyba and H. Sun; J.S. Mandapaka and K. Namuduri; and W. Wang and F. Fund. The AERPAW platform development and field experiment execution were carried out by Ö. Özdemir, A. Gürses, İ. Güvenç, M.L. Sichitiu, R. Dutta, M. Mushi, and T. Zajkowski.

%\cite{gurses2024digital}.
% \\
% \text{\hspace{1em}} F.A., S.A., and T.A. curated and analysed the data, and wrote parts of the manuscript.
% S.A. reviewed the curation and wrote parts of the manuscript. All authors reviewed the manuscript.
% \\
% \text{\hspace{1em}} The article authors have declared no conflicts of interest.
\bibliographystyle{IEEEtran}
% \section*{REFERENCES}
\bibliography{mybib}
\end{document}